\documentclass[aps,pre,nofootinbib,showpacs,twocolumn]{revtex4-1}%
\usepackage{amssymb,latexsym,mathrsfs}
\usepackage{amsfonts}
\usepackage{mathrsfs}
\usepackage{amsmath}
\usepackage{graphicx}
\usepackage{color}

\newcommand{\beq}{\begin{equation}}
\newcommand{\eeq}{\end{equation}}
\newcommand{\beqn}{\begin{eqnarray}}
\newcommand{\eeqn}{\end{eqnarray}}
\newcommand{\bearr}{\begin{array}}
\newcommand{\enarr}{\end{array}}

\def\bea{\begin{eqnarray}}
\def\eea{\end{eqnarray}}
\def\ba{\begin{array}}
\def\ea{\end{array}}

\begin{document}
\title{Sensitivity to initial conditions of a $d$-dimensional long-range-interacting quartic Fermi-Pasta-Ulam model: Universal scaling}
\author{Debarshee  Bagchi$^1$}
\email[E-mail address: ]{debarshee@cbpf.br}
\author{Constantino Tsallis$^{1,2}$}
\email[E-mail address: ]{tsallis@cbpf.br}
\affiliation{$^1$Centro Brasileiro de Pesquisas Fisicas  \\ and National Institute of Science and Technology of Complex Systems, Rua Xavier Sigaud 150, 22290-180 Rio de Janeiro-RJ,  Brazil \\
and \\
$^2$Santa Fe Institute, 1399 Hyde Park Road, New Mexico 87501, USA }
\date{\today}

\begin{abstract}
We introduce a generalized  $d$-dimensional Fermi-Pasta-Ulam (FPU) model in presence of  long-range
interactions, and perform a first-principle study of its chaos for $d=1,2,3$ through large-scale numerical simulations. The
nonlinear interaction is assumed to decay algebraically as $d_{ij}^{-\alpha}$  ($\alpha \ge 0$), $\{d_{ij}\}$ being the
distances between $N$ oscillator sites. Starting from random initial conditions we compute the maximal Lyapunov
exponent $\lambda_{max}$ as a function of $N$. Our $N>>1$ results strongly indicate that $\lambda_{max}$
remains constant and positive for $\alpha/d>1$ (implying strong chaos, mixing and ergodicity), and that it vanishes
like $N^{-\kappa}$ for $0 \le \alpha/d < 1$ (thus approaching weak chaos and opening the possibility of breakdown
of ergodicity).  The suitably rescaled exponent $\kappa$ exhibits universal scaling, namely that $(d+2) \kappa$ depends
only on $\alpha/d$ and, when $\alpha/d$ increases from zero to unity, it monotonically decreases from unity to zero,
remaining so for all $\alpha/d >1$. The value $\alpha/d=1$ can therefore be seen as a critical point separating the ergodic
regime from the anomalous one, $\kappa$ playing a role analogous to that of an order parameter. This scaling law is
consistent with Boltzmann-Gibbs statistics for $\alpha/d > 1$, and possibly with $q$-statistics for $0 \le \alpha/d < 1$.
\end{abstract}

\pacs{05.70.-a, 05.45.Pq, 05.45.-a, 89.75.Da}

\maketitle


\section{Introduction}

Many-body systems with long-range-interacting forces are very important in nature, the primary example being gravitation.
Long-ranged systems deviate significantly from the conventional `well behaved' systems in many respects. Various features 
like ergodicity breakdown, ensemble inequivalence, non-mixing nonlinear dynamics, partial (possibly hierarchical) occupancy 
of phase space, thermodynamical nonextensivity for the total energy, longstanding metastable states, phase transitions even 
in one dimension, and other anomalies, can be observed in systems with long-range interactions. Consistently, some of the 
usual premises of Boltzmann-Gibbs (BG) statistical mechanics are challenged and an alternative thermostatistical description 
of these systems becomes necessary in many instances. For some decades now, $q$-statistics \cite{qstat, qstat2} has been 
a useful formalism to study such systems, and has led to satisfactory experimental validations for a wide variety 
of complex systems (see for instance \cite{cs1,cs2,cs3,cs4,cs5,cs6,cs7,cs8,cs9,cs10,cs11,cs12,cs13,cs14,cs15}). The deep 
understanding of the microscopical nonlinear dynamics of such systems naturally constitutes a must in order to theoretically 
legitimize the efficiency of the $q$-generalization of the BG theory. For classical systems such as many-body Hamiltonian ones 
and low-dimensional maps, a crucial aspect concerns the sensitivity to the initial conditions, which is characterized by the spectrum 
of Lyapunov exponents. If the maximal Lyapunov exponent $\lambda_{max}$ is positive, mixing and ergodicity are essentially 
warranted, and we consequently expect the BG entropy and statistical mechanics to be applicable. If instead $\lambda_{max}$ 
vanishes, the sensitivity to the initial conditions is subexponential, typically a power-law with time, and we might expect nonadditive 
entropies such as $S_q$ and its associated statistical mechanics to emerge, as has been observed numerically as well as experimentally
in many systems (see, for instance, \cite{LyraTsallis, AnteneodoTsallis, BaldovinRobledo,Casati, TsallisAnanos, cs15, cs14,TirnakliBorges}).


\section{Model and the numerical scheme}
In the present paper we extend to $d$-dimensions ($d=1,2,3$) and numerically study from first principles (i.e., using only Newton's 
law $\vec F=m \vec a$) the celebrated Fermi-Pasta-Ulam (FPU) model with periodic boundary conditions; nonlinear long-range 
interactions between all the $N=L^d$ oscillators are allowed as well. The Hamiltonian is the following one:
\begin{equation}
\mathcal{H} = \sum_{i} \frac {\vec p_i\,^2}{2 m_i}  + \frac a {2} \sum_{i} (\vec r_{i+1} - \vec r_i)^2  
+ \frac {b} {4 \tilde N} \sum_{i} \sum_{j \ne i} \frac {(\vec r_i - \vec r_j)^4} {d_{ij}^{~\alpha}}
\label{H}
\end{equation}
where $\vec r_i$ and $\vec p_i$ are the displacement and momentum of the $i$-th particle with mass $m_i \equiv m$; $a \ge 0$, $b>0$, and $\alpha \ge 0$. Here $d_{ij}$ 
is the shortest Euclidean distance between the $i$-th and $j$-th lattice sites ($1 \le i,j \le N$); this distance depends on the geometry of the lattice (ring, periodic square
or cubic lattices). Thus for $d = 1$,  $d_{ij} = 1,2,3, ...$; for $d = 2$, $d_{ij} = 1, \sqrt 2, 2, ...$, and, for $d = 3$, $d_{ij} = 1, \sqrt 2, \sqrt 3, 2, ...$ If $\alpha/d >1$ ($0 \le \alpha/d \le 1$) 
we have short-range (long-range) interactions in the sense that the potential energy per particle converges (diverges) in the thermodynamic limit $N\to\infty$; in particular, the 
$\alpha \to\infty$ limit corresponds to only first-neighbor interactions, and the $\alpha=0$ value corresponds to typical mean field approaches, when the coupling constant is 
assumed to be independent from distance.  The instance $(d,\alpha)=(1,\infty)$ recovers the original $\beta$-FPU Hamiltonian, that has been profusely studied in the literature; 
the $d=1$ model and generic $\alpha$ has been addressed in \cite{1DFPU}.

Although not necessary (see \cite{AnteneodoTsallis}), we have followed the current use and have made the Hamiltonian extensive for all values of $\alpha/d$ by adopting the 
scaling factor $\tilde N$ in the quartic coupling, where
\begin{equation}
\tilde N \equiv \sum_{i =1}^{N} \frac{1}{d_{ij}^{~\alpha}}
\label{Ntilde}
\end{equation}
hence, $\tilde N$ depends on $\alpha, N, d$, and the geometry of the lattice. Note that for $\alpha = 0$ we have $\tilde N = N$, 
which recovers the rescaling usually introduced in mean field approaches. In the thermodynamic limit $N \to\infty$, $\tilde N$
remains constant for $\alpha/d >1$, whereas $\tilde N \sim \frac{N^{1-\alpha/d}}{1-\alpha/d}$ for $0 \le \alpha/d <1$ 
($\tilde N \sim \ln N$ for $\alpha/d=1$); see details in \cite{AnteneodoTsallis} and references therein.

Let us mention that the analytical thermostatistical approach of the present model is in some sense even harder than that of 
coupled XY or Heisenberg rotators already addressed in \cite{AnteneodoTsallis,XY_uc,XYmodel,CirtoTsallis2014,CirtoNobre2015}. 
Indeed, the standard BG approach of these models is analytically tractable, whereas not even that appears to be possible for the 
original FPU, not to say anything for the present generalization. Therefore, for this kind of many-body Hamiltonians, the numerical 
approach appears to be the only tractable one.

To numerically solve the equations of motion (Newton's law)
we have employed the symplectic second order accurate velocity Verlet algorithm. 
To accelerate the computationally expensive part of the force calculation routine we have exploited the convolution theorem and 
used a Fast Fourier transform algorithm. This yields a considerable reduction in the number of operations for force calculation from
$\mathcal{O}(N^2)$ to $\mathcal{O}(N\ln N)$, thus facilitating computation for  larger system sizes and longer times. 

We choose the time step $\Delta t$ (which is typically $ \sim 10^{-3}$ for most of our results) such that the standard deviation of the energy 
density over the entire simulation time (i.e., the number of iterations required by the maximal Lyapunov exponent to saturate, which is typically 
$\sim 10^5-10^6$ iterations, depending on system parameters) is of the order of $10^{-4}$ or smaller (for the range of $N$ considered 
here, $10 < N < 10^6$).

Starting from a random initial displacements $\vec r_i$ drawn from a uniform distribution centered around zero, and momenta $\vec p_i$
from a Gaussian distribution with zero mean and unit variance, we evolve the system and compute the maximal Lyapunov exponent 
$\lambda_{max}$ defined as follows: 
\begin{equation}
\lambda_{max} =\lim_{t \to\infty} \lim_{\delta(0) \to 0} \frac 1 t \ln \frac {\delta(t)}{\delta(0)} \,,
\end{equation}
where $\delta(t) = \sum_i (\delta r_i^2 + \delta p_i^2)^{1/2}$ is the metric distance between the fiducial orbit and the reference orbit having initial 
displacement $\delta(0)$. We numerically compute this quantity by using the algorithm by Benettin et al \cite{Benettin}. For typical values of the 
exponent $\alpha$, we compute $\lambda_{max}$ as a function of the system size $N$ for $d = 1,2$, and $3$. 


\section{Simulation Results}
Let us now present the results of our numerical analysis by setting $m=1$  (no loss of generality), and fixing the energy density $u \equiv U/N= 9.0$ and
$b = 10.0$ for all $d$, unless stated otherwise, where $U$ is the total energy associated with $\mathcal {H}$. Additionally, we have set the harmonic term
to zero, i.e. $a=0$, for reasons that will be elaborated later. In fact such a model, with only the quartic anharmonic nearest neighbor interactions, has been
studied previously in the context of heat conduction \cite{hcond}.
\begin{figure}[htb]
\centerline
{\includegraphics[width=6.0cm,angle=-90]{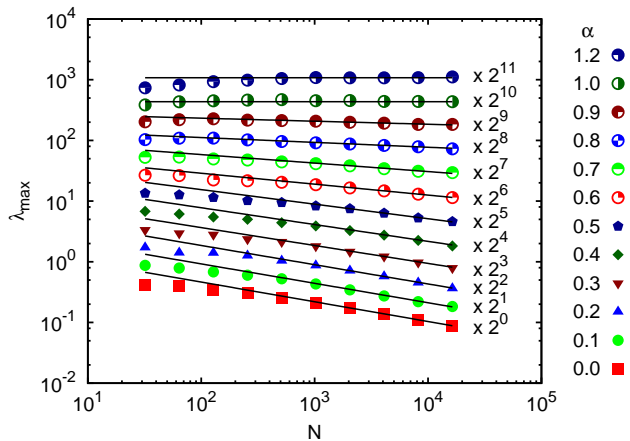}}
\caption{(Color online) Log-log plot of the dependence, for a ring ($d = 1$), of the maximal Lyapunov exponent $\lambda_{max}$  on the number $N=L$ of 
oscillators for $(a,b,u)=(0,10,9)$ and typical values of the exponent $\alpha$. Each individual curve has been multiplied by the number indicated next to it 
for visualization clarity.}
\label{fig:FPU1Da0}
\end{figure}

\begin{figure}[htb]
\centerline
{\includegraphics[width=6.0cm,angle=-90]{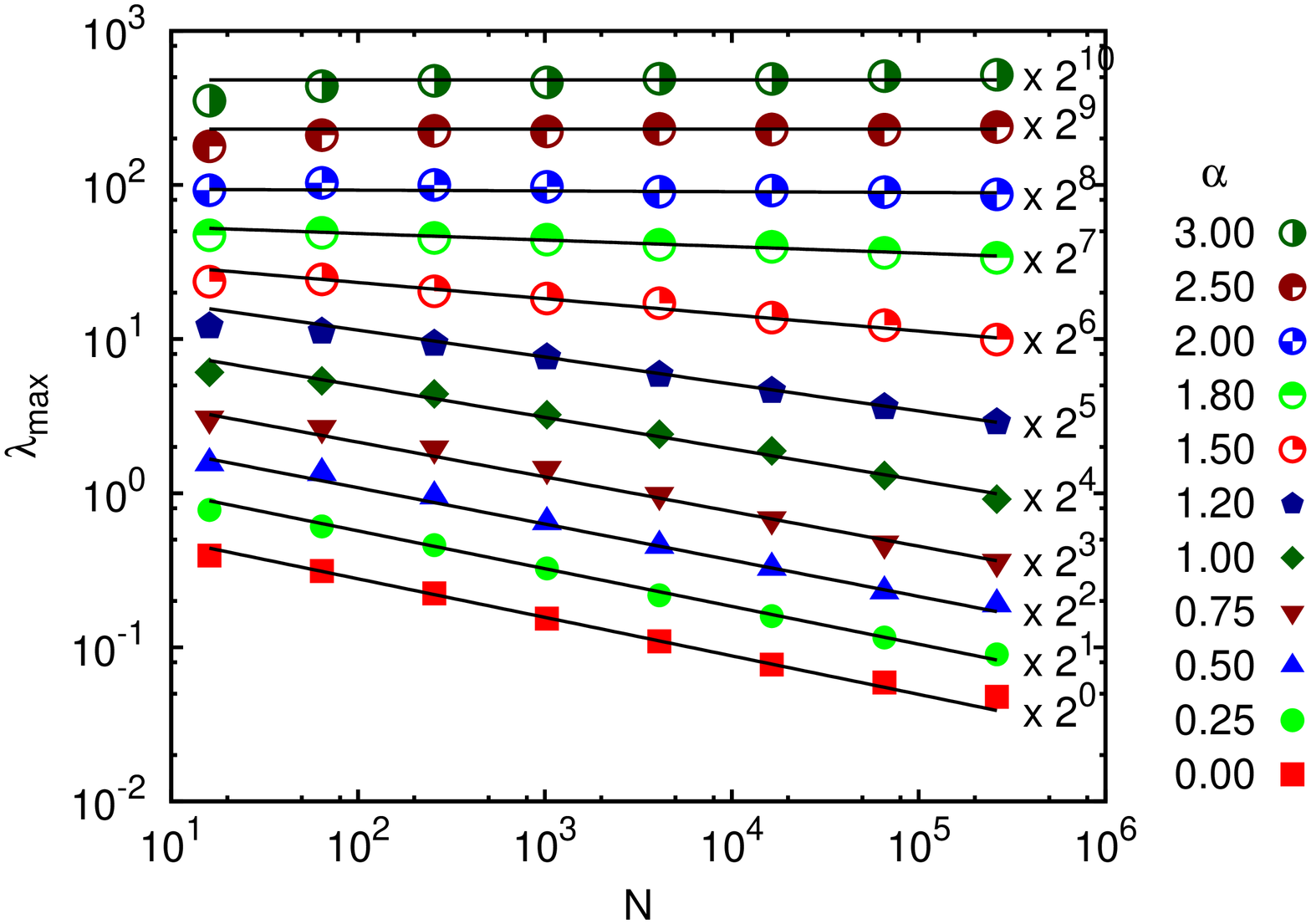}}
\caption{(Color online) The same as in Fig. 1 for a periodic square lattice ($d=2$) with $N=L^2$ oscillators.}
\label{fig:FPU2Da0}
\end{figure}

\begin{figure}[htb]
\centerline
{\includegraphics[width=6.0cm,angle=-90]{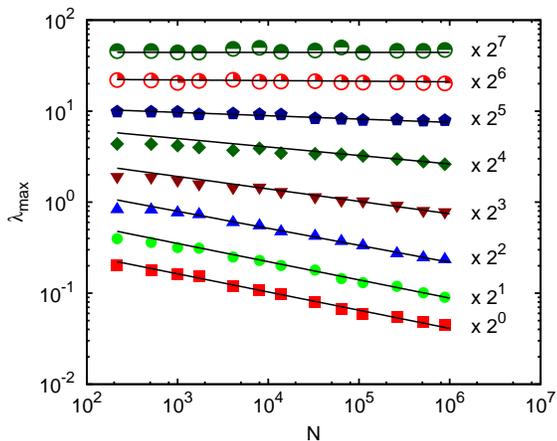}}
\caption{(Color online) The same as in Fig. 1 for a periodic cubic lattice ($d=3$) with $N=L^3$ oscillators.}
\label{fig:FPU3Da0}
\end{figure}

In Figs. \ref{fig:FPU1Da0},  \ref{fig:FPU2Da0} and \ref{fig:FPU3Da0} we present, for $d = $1, 2 and 3 respectively, the maximal Lyapunov exponent $\lambda_{max}$ 
as a function of the system size for typical values of the exponent $\alpha$. We find that, for $\alpha > d$, $\lambda_{max}$ saturates to a positive value with increasing
$N$, which strongly suggests that it will remain so for $N \to\infty$, thus leading to ergodicity, which in turn legitimizes the BG thermostatistical theory. In  contrast, 
for $0 \le \alpha < d$, $\lambda_{max}$ algebraically decays with $N = L^d$ as 
\begin{equation}
\lambda_{max} \sim N^{-\kappa}
\end{equation}
where $\kappa > 0$ and depends on $(\alpha,d)$. Assuming that it remains so for increasingly large $N$, we expect $\lim_{N\to\infty}  \lambda_{max} =0$,
which implies that the entire Lyapunov spectrum vanishes. This characterizes weak chaos for $0 < \alpha /d < 1$, i.e., subexponential sensitivity to the
initial conditions, which opens the door for breakdown of mixing, or of ergodicity, or some other nonlinear dynamical anomaly. Within this scenario, the 
violation of Boltzmann-Gibbs statistical mechanics in the $N \to\infty$ limit becomes strongly plausible (see, for example, \cite{TirnakliBorges,1DFPU}).


\begin{figure}[htb]
{\includegraphics[width=6.0cm,angle=-90]{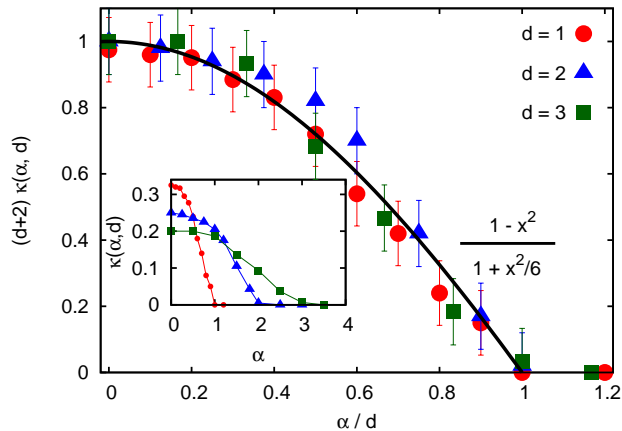}}
\caption{(Color online) The inset shows the exponent $\kappa(\alpha,d)$ as a function of $\alpha$ for $d = 1,2,3$. Note that 
$\kappa > 0$ for $0 \le \alpha < d$ and $\kappa = 0$ for $\alpha > d$.  The main figure exhibits the universal law obtained by appropriately rescaling 
the abscissas and ordinates as indicated on the axes, i.e., $(d+2) \kappa(\alpha,d) =f(\alpha/d)$.  The thick continuous curve is the heuristic 
scaling function $f(x) = (1-x^2)/(1+x^2/6)$ \cite{XYmodel}, which, within the present precision, is a remarkably close fit to the collapsed data. 
The present collapse obviously implies $\kappa(0,d)=1/(d+2)$, hence $\lim_{d \to\infty} \kappa(0,d)=0$, thus recovering ergodicity, as intuitively expected.
}
\label{fig:kappa_a_mul}
\end{figure}

From the results illustrated in Figs. \ref{fig:FPU1Da0}, \ref{fig:FPU2Da0} and \ref{fig:FPU3Da0} we compute the exponent $\kappa(\alpha,d)$ for
$d = 1,2$ and $3$, as shown in Fig. \ref{fig:kappa_a_mul}, including its inset.
We find that $\kappa(\alpha,d) > 0$ for $0 \le \alpha < d$,  and, within numerical accuracy, vanishes for $\alpha > d$.
Also note that $\kappa(0,d)$ decreases for increasing $d$. Remarkably enough, all three curves
in the inset of Fig. \ref{fig:kappa_a_mul} can be made to collapse onto a single curve through the  
scalings $\alpha \to \alpha/d$ and $\kappa(\alpha,d) \to (d+2)~\kappa(\alpha,d)$. This is shown in the main figure of 
Fig. \ref{fig:kappa_a_mul}. {\it In other words, $(d+2)~\kappa(\alpha,d) = f(\alpha/d)$ where $f(x)$ appears to be a universal function.}

A similar scaling was also verified for the classical model of long-ranged coupled rotators \cite{AnteneodoTsallis,XYmodel}. Some relevant differences exist 
however between the two models and their sensitivities to initial conditions. The long-range-interacting planar rotator model exhibits, for a critical energy density 
$u_c$ \cite{AnteneodoTsallis,XY_uc,XYmodel,CirtoTsallis2014}, a second order phase transition from a clustered phase (ferromagnetic) to a homogeneous one (paramagnetic).  
Such critical phenomenon does not exist in either the short-ranged or the long-ranged FPU model. For the XY ferromagnetic model the exponent $\kappa$ for $\alpha = 0$ is found 
to be independent from $d$ (quite obvious since the $\alpha=0$ model has no dimension) and given by $\kappa(0,d) = 1/3$ \cite{Firpo,XYmodel} (see also \cite{Bachelard}). 
In contrast, our long-range model yields a value $\kappa(0,d)$ which depends on $d$. Indeed, for $d=$ 1, 2 and 3, we respectively obtain $\kappa(0,d) \simeq$ 1/3, 1/4 and 1/5.

This difference in $\kappa(0,d)$ is related to the fact that, for the XY model, the number of degrees of freedom (number of independent variables needed to
specify the state of the system in phase space) for $N$ coupled rotators in $d$ dimensions is $2 N \,(\forall d)$, whereas, for our model,
there are $2Nd$ degrees of freedom, hence the dimension of the full phase space grows linearly with $d$. Thus there are more possible phase
space dimensions for our coupled oscillator system to escape even if gets somewhat trapped in some non-chaotic region of the phase space. 
Consequently, the system gets closer to ergodicity (equivalently, $\kappa$ gets closer to zero) for increasing $d$. It is even not excluded that, 
because of some generic reason of this kind, $\kappa(0,d) \,(\forall d)$ for the long-ranged XY model  and  $\kappa(0,1)$ for the system studied
here, we obtain (in absence of the integrable term, i.e., with $a = 0$) the same value $1/3$. 

In this context we should mention another recent study \cite{HMFmodel} of the Hamiltonian mean field (HMF) model which is  the  $\alpha = 0$ particular 
case of the long-ranged XY model discussed above. Using numerical and analytical arguments it was suggested that the nature of chaos is quite different 
for this model (which has a phase transition at $u_c = 3/4$) in the homogeneous phase ($u > u_c$)  where $\lambda_{max} \sim N^{-1/3}$, the ordered
phase ($u < u_c$) where $\lambda_{max}$ remains positive and finite, and at criticality ($u \to u_c$) where $\lambda_{max} \sim N^{-1/6}$ in the infinite 
size limit.
However in another earlier work \cite{one-ninth-exp}, using scaling arguments and numerical simulations, it was observed that $\lambda_{max} \sim N^{-1/9}$ 
below the critical point ($u = 0.69$) in the (non equilibrium) quasi-stationary regime of the HMF system.

Another class of models might also have a similar behavior. If we consider the $d$-dimensional long-range-interacting $n$-vector ferromagnet, 
we expect an exponent $\kappa(\alpha,n,d)$. We know that for $n=2$ (XY symmetry) $\kappa(0,2,1)=1/3$, for $n=3$ (classical Heisenberg 
model symmetry) $\kappa(0,3,1) = 0.225 \pm 0.030$ \cite{NobreTsallis}, and for $n \to\infty$ (spherical model symmetry) most plausibly 
$\kappa(0,\infty,d) = 0\ (\forall d)$. These expressions can be simply unified through $\kappa(0,n,d)=1/(n+1)\ (\forall d)$.

Strikingly enough, the present Fig. \ref{fig:kappa_a_mul}  and Fig. 2 of \cite{XYmodel} for the $d$-dimensional XY model are numerically 
indistinguishable within error bars. This suggests the following heuristic expression:
\begin{equation}
\frac{\kappa(\alpha,d)}{\kappa(0,d)} =  f(\alpha/d) \simeq \frac{1-(\alpha/d)^2}{1+(\alpha/d)^2/6} \,,
\end{equation}
where this specific analytic expression for $f(x)$ has been first suggested in \cite{XYmodel}. This or a similar universal behavior is expected to hold 
for $d$-dimensional long-range-interacting many-body models such as the present one, the XY ferromagnetic one, and others such as, for instance,
the $n$-vector ferromagnetic one ($\forall n$).
 
\begin{figure*}[htb]
{
\centering
\includegraphics[width=4.25cm,angle=-90]{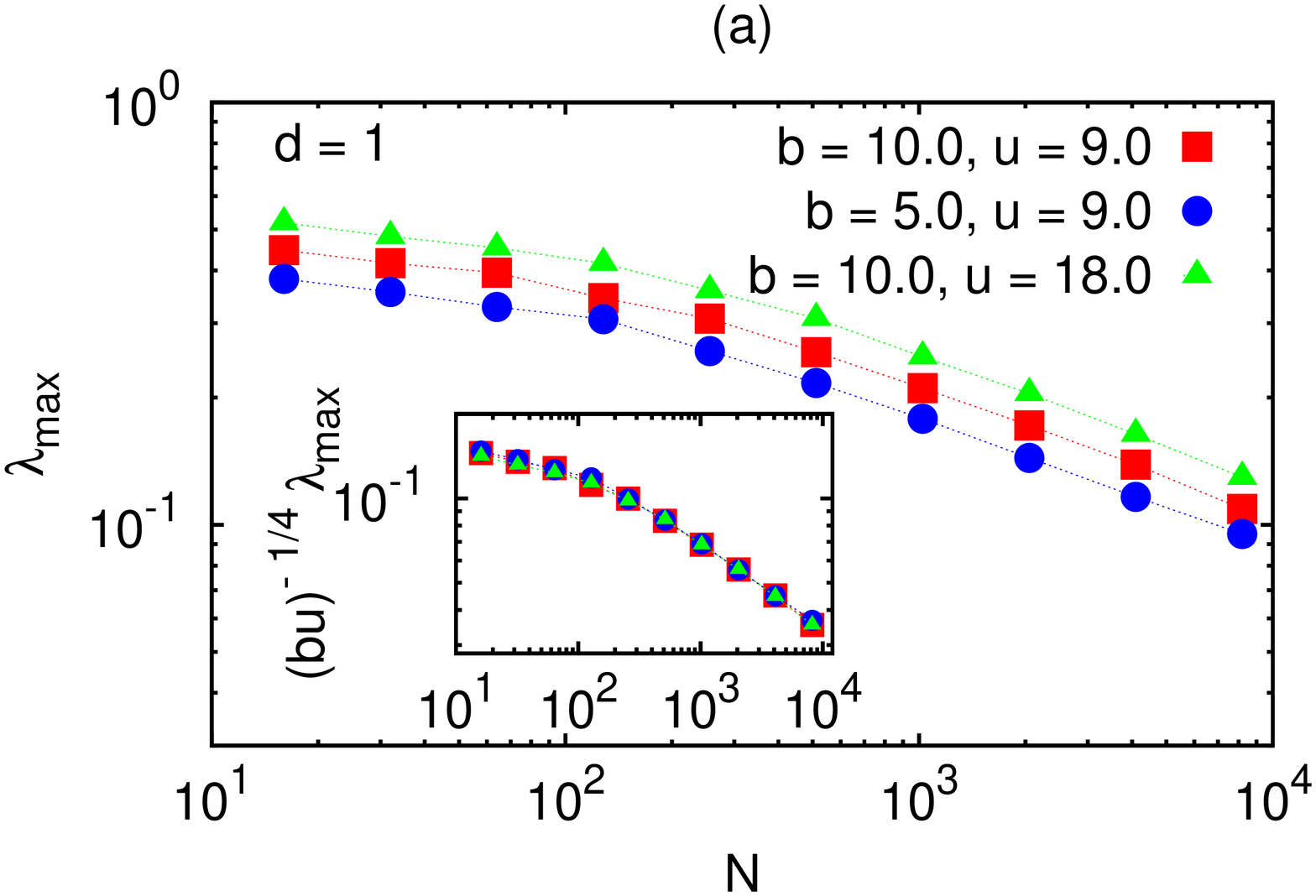} \hskip-0.3cm
\includegraphics[width=4.25cm,angle=-90]{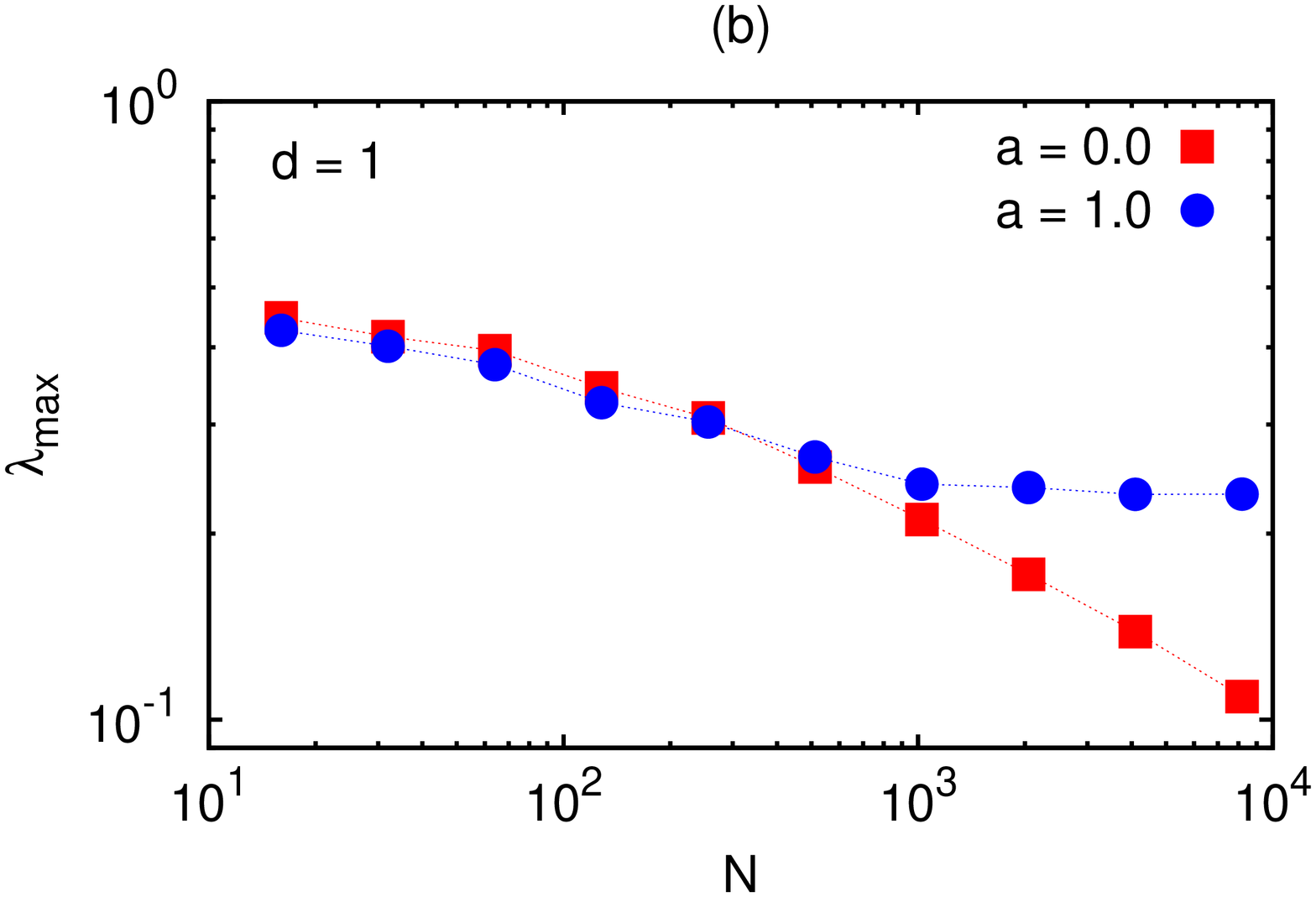}  \hskip-0.3cm
\includegraphics[width=4.25cm,angle=-90]{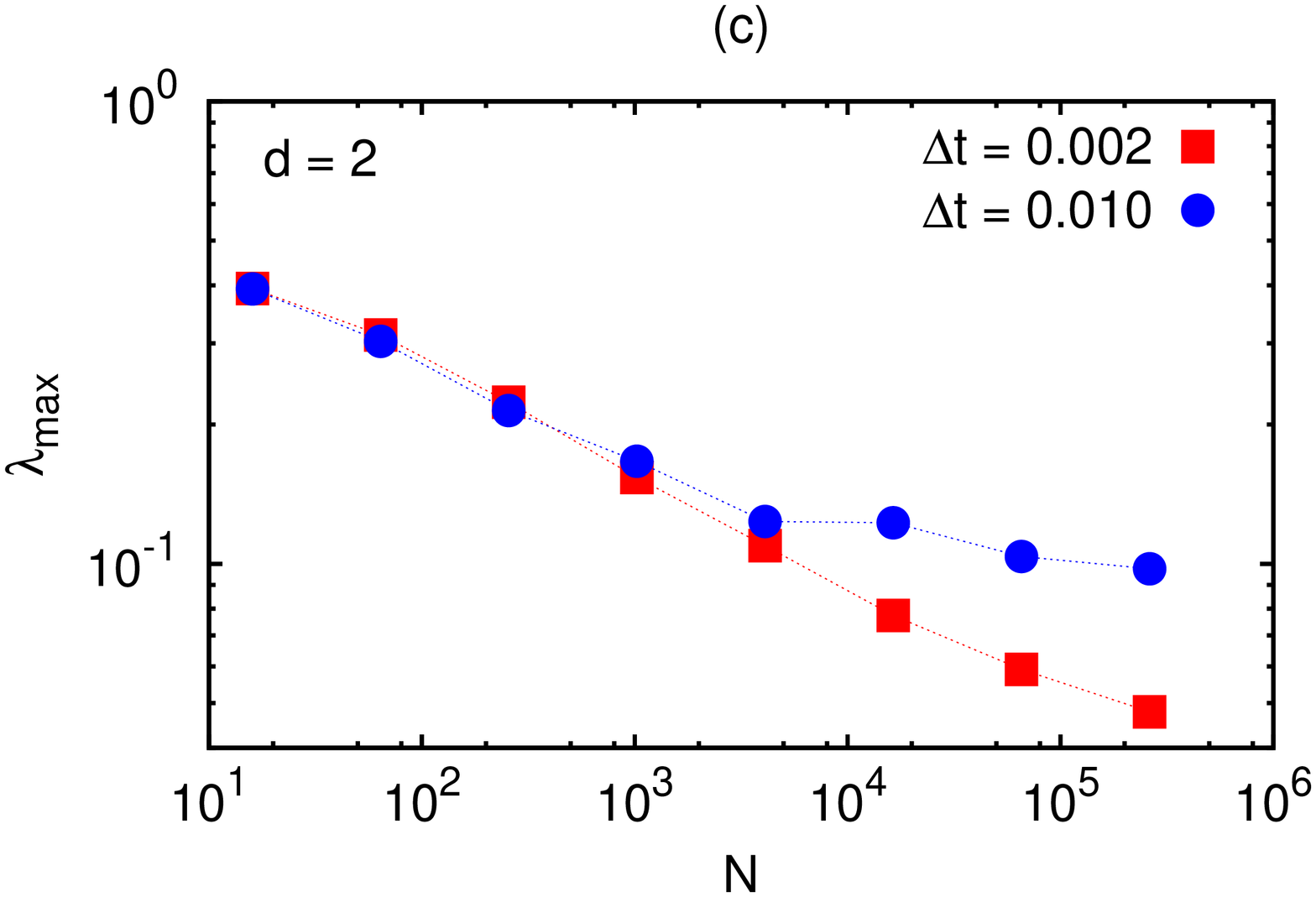}
}
\caption{(Color online) Parameter dependencies of $\lambda_{max}(N)$ with $N=L^d$: (a) for different $b$'s and $u$'s with  ($a,\alpha,\Delta t$) = $(0,0,0.002)$ 
-- inset shows data collapse obtained by rescaling the y-axis of the main figure as $(bu)^{-1/4} ~ \lambda_{max}$; (b) for different $a$'s with ($\alpha,b,u,\Delta t$) = $(0,10,9, 0.002)$;
(c) for different $\Delta t$'s with ($a,\alpha,b,u$) = $(0,0,10,9)$.}
\label{fig:parameters}
\end{figure*}

All the numerical results presented until now are with a fixed set of parameters $(a,b,u,)=(0,10,9)$ and a fixed time step $\Delta t$.
Before concluding, let us briefly mention some results concerning the influence of these parameters on $\lambda_{max}(N)$ and $\kappa(\alpha,d)$.
In Fig. \ref{fig:parameters}a we plot  $\lambda_{max} (N)$ for $d = 1$ for three different sets of $(b,u)$ keeping all other parameters unchanged.
We find that increasing $b$ has the same effect as increasing $u$ -- the maximum Lyapunov exponent $\lambda_{max}$ increases with both of them but the slope of the curve
$\kappa$ remains practically unaltered.
For $a = 0$, it is straightforward to show that the average of Hamiltonian Eq. (\ref{H}) remains invariant with respect to $b$ and $u$ (all other
parameters remaining the same) under the transformations
 
\begin{equation}
x ^\prime = (b/u)^{1/4} ~ x \,, ~~~~~~~~~~~
t ^\prime = (bu)^{1/4} ~t.
\label{transform}
\end{equation}
 
The second transformation in Eq. (\ref{transform}) implies that the maximum Lyapunov exponent $\lambda_{max}$ ($\sim t^{-1}$) satisfies the following scaling relation:

\begin{equation}
\lambda ^\prime_{max} = (bu)^{-1/4} ~ \lambda_{max} \,.
\label{scaling}
\end{equation}

Using the data in the main figure, we show in the inset of Fig. \ref{fig:parameters}a the variation of $\lambda ^\prime_{max} \equiv (bu)^{-1/4} ~ \lambda_{max}$
with $N$. As predicted by the scaling analysis, we get an excellent data collapse of the three curves.
This is precisely as desired, keeping in mind the universal behavior ubiquitously found in statistical mechanics, in the sense that scaling
indices, such as $\kappa$ here, are generically expected to be independent of the microscopic details of the model.

For nonzero values of $a$, the simple scaling Eq. (\ref{scaling}) disappears, and $\lambda_{max}(N)$ shows a saturation to a positive value that vanishes
for $a = 0$ when $N$ is large, $b$ being a finite positive number. This is shown in Fig. \ref{fig:parameters}b for two values of $a$ with the same value of $b$. 
The saturation of $\lambda_{max}$ for $a > 0$ needs careful study to be understood properly. In Fig. \ref{fig:parameters}c we have shown (for $d = 2$) that
increasing $\Delta t$ can also lead to a deviation from the $\lambda_{max} \sim N^{-\kappa}$ behavior; this deviation is quite expected, and one should choose the
time step judiciously. Note that the saturation behavior in Fig. \ref{fig:parameters}b is not due to finiteness of the time step.


\section{Summary and discussions}
Summarizing, we have introduced a $d$-dimensional generalization of the celebrated Fermi-Pasta-Ulam model which allows for long-range nonlinear 
interaction between the oscillators, whose coupling constant decays as  $distance^{-\alpha}$. We have then focused on the sensitivity to initial conditions, more 
precisely on the first-principle (based on Newton's law) calculation of the maximal Lyapunov exponent $\lambda_{max}$ as a function of the number $N$ of 
oscillators using large-scale numerical simulations. Without the quadratic nearest neighbor interaction (i.e., $a=0$),  $\lambda_{max}(N)$ appears to asymptotically behave as 
$N^{-\kappa}$ (with $\kappa >0$) for $0 \le \alpha/d <1$, and approach a positive constant (i.e., $\kappa=0$) for $\alpha/d>1$ in the $N\to\infty$ thermodynamic 
limit. Our results provide strong indication that $\kappa$ only depends on $(\alpha,d)$, and does so in a universal manner, namely $(2+d)\kappa(\alpha,d)=f(\alpha/d)$ for
$0 \le \alpha/d <1$, and $\kappa=0$ for $\alpha/d >1$. This universal suppression of strong chaos is well approximated by a model-independent heuristic function
$f(x)\simeq (1-x^2)/(1+x^2/6)$, previously found \cite{AnteneodoTsallis,XYmodel} for the $d$-dimensional XY model of coupled rotators. 
Thus, in the thermodynamic limit, these systems (and plausibly others as well) have a sort of critical point at $\alpha/d=1$, which separates the ergodic 
$\alpha/d>1$ region (where the Boltzmann-Gibbs statistical mechanics is valid, and  the stationary state distribution of velocities is the standard Maxwellian one), from the weakly 
chaotic $0 \le \alpha/d < 1$ region with anomalous nonlinear dynamical behavior (where $q$-statistics might be expected to be valid, and the one-body distribution of 
velocities appears to be of the $q$-Gaussian form, consistently with preliminary results available in the literature \cite{1DFPU,CirtoTsallis2014}). The present universality 
scaling for $\kappa(\alpha/d)$ enables the conjecture that the indices $q$ of the distributions of velocities and of energies might exist and only depend on the ratio 
$\alpha/d$. Naturally, all these observations need further and  wider checking, which would be welcome. Work along this line is in progress.\\

{\bf Acknowledgments:} We are thankful to H. Christodoulidi and A. Ponno for sharing with us their observations concerning the role of the coupling constant $a$ in the
determination of $\kappa$. We have also benefitted from  fruitful discussions with L.J.L. Cirto, G. Sicuro, P. Rapcan, T. Bountis, and E.M.F. Curado. We gratefully 
acknowledge partial financial support from CNPq and Faperj (Brazilian agencies) and the John Templeton Foundation-USA.

\end{document}